\documentclass[twocolumn,aps,prd,amsmath,amsfonts,nofootinbib]{revtex4-1}

\newcommand{\ud}{\mathrm{d}}

\newcommand{\bs}{\begin{split}}
\newcommand{\es}{\end{split}}

\usepackage{xcolor}
\usepackage{verbatim}
\usepackage{float}
\usepackage{color}
\usepackage[hyperindex,breaklinks]{hyperref}
\hypersetup{
colorlinks,
    linkcolor={black},
    citecolor={blue!50!black},
    urlcolor={blue!80!black},
    pdftitle={Lambda does not Lens},
    pdfauthor={Luke M. Butcher}}
\usepackage{slashed}
\usepackage{graphicx}
\usepackage{subfigure}
\usepackage{psfrag}
\usepackage{mathtools}
\usepackage{amsthm}
\theoremstyle{plain}

\begin{document}

\title{Lambda does not lens:\\ Deflection of light in the Schwarzschild--de Sitter spacetime}
\author{Luke M. Butcher}
\email[]{lmb@roe.ac.uk}
\affiliation{Institute for Astronomy, University of Edinburgh, Royal Observatory, Edinburgh EH9 3HJ, United Kingdom}
\date{June 16, 2016}

\begin{abstract}
Debate persists as to whether the cosmological constant $\Lambda$ can directly modify the power of a gravitational lens. With the aim of reestablishing a consensus on this issue, I conduct a comprehensive analysis of gravitational lensing in the Schwarzschild--de Sitter spacetime, wherein the effects of $\Lambda$ should be most apparent. The effective lensing law is found to be in precise agreement with the $\Lambda=0$ result: $\alpha_\mathrm{eff} = 4m/b_\mathrm{eff}+15\pi m^2/4b_\mathrm{eff}^2 +O(m^3/b_\mathrm{eff}^3)$, where the effective bending angle $\alpha_\mathrm{eff}$ and impact parameter $b_\mathrm{eff}$ are defined by the angles and angular diameter distances measured by a comoving cosmological observer. [These observers follow the timelike geodesic congruence which (i) respects the continuous symmetries of the spacetime and (ii) approaches local isotropy most rapidly at large distance from the lens.] The effective lensing law can be derived using lensed or unlensed angular diameter distances, although the inherent ambiguity of unlensed distances generates an additional uncertainty $O(m^5/\Lambda b_\mathrm{eff}^7)$. I conclude that the cosmological constant does not interfere with the standard gravitational lensing formalism.
\end{abstract}
\maketitle

\section{Introduction}\label{intro}
At present there is no clear consensus regarding the effect of the cosmological constant $\Lambda$ on gravitational lensing. For many years, the prevailing view \cite{Islam83,Finelli07} held that $\Lambda$ would not modify the standard lensing formalism: all cosmological effects were already accounted for within angular diameter distances. 
This belief stems from a simple observation: $\Lambda$ does not appear in the orbital equation of light in the Schwarzschild--de Sitter (SdS) spacetime. Thus, outside a static spherically symmetric gravitational lens, light follows a path which is \emph{independent} of the cosmological constant. However, Rindler and Ishak have recently identified two key flaws in this long-accepted argument \cite{Rindler07}. First, the SdS spacetime is not asymptotically flat, so the standard treatment of gravitational lensing (using angles and distances defined at infinity) no longer applies. Second, although the lightpath is independent of $\Lambda$, the \emph{physical angles} formed by this path are quantified by the SdS metric, which \emph{does} depend on $\Lambda$. Their calculations reveal a $\Lambda$-dependent contribution to a particular physical angle, suggesting the existence of a previously unnoticed $\Lambda$--lensing interaction that could significantly weaken the observed lensing power of massive bodies at high redshift. Unfortunately their analysis does not proceed far enough to be definitive: the motion of cosmological observers is ignored, and there is no attempt to determine whether or not angular diameter distances already account for the effect. Consequently, a substantial controversy has arisen over the existence of this $\Lambda$--lensing interaction, with many authors building on or corroborating Rindler and Ishak's original results \cite{Ishak08a,Ishak08b, Sereno08,Sereno09,Schucker09a,Schucker09b}, while others advance new arguments for the conventional wisdom \cite{Khrip08,Park08,Simpson10}; see \cite{Ishak2010} for a review.

In this paper I hope to re-establish a consensus on this issue by revisiting the spacetime that first sparked the controversy: Schwarzschild--de Sitter. The aim is to complete the rigorous analysis that Rindler and Ishak began, and hence establish whether or not $\Lambda$ actually alters the observable effects of gravitational lensing. The key features of the analysis will be as follows:
\begin{itemize}
\item We will focus on the Schwarzschild--de Sitter spacetime, which represents a point-mass gravitational lens embedded within a de Sitter cosmology. This starkly simple scenario is such an extreme case (as compared to, say, a spatially dispersed mass in a more realistic $\Lambda$CDM background) that were a $\Lambda$--lensing effect to exist, one would certainly expect it to manifest itself here. As such, the Schwarzschild--de Sitter spacetime serves as a \emph{litmus test} for the existence of a $\Lambda$--lensing interaction in general. 
\item The derivations will be thorough and self-contained, working from first principles where appropriate. Heeding Rindler and Ishak, we will address the lack of asymptotic flatness and the metric's role in defining physical angles. In contrast to \cite{Rindler07}, however, we will also account for the motion of cosmological observers and the effect of $\Lambda$ on angular diameter distances. 
\item Care will be taken in tracking the magnitude of neglected quantities; this includes the uncertainties inherent in unlensed angular diameter distances, which have been ignored elsewhere.
\item The physical predictions of our calculations will be summarized in terms of an \emph{effective} lensing law. To this end, we will formulate effective definitions of the bending angle and impact parameter, extending the meaning to these concepts beyond asymptotically flat spaces.
\end{itemize}
It will also be expedient to assume that the light source, the lens, and the observer are precisely collinear; we will generalise our results beyond this assumption at the end of the calculation. We work in units such that $c=G=1$.

\section{Schwarzschild--de Sitter}
We begin with Kottler's metric for the Schwarzschild--de Sitter (SdS) spacetime \cite{Kottler18}:
\begin{align}\nonumber
\ud s^2 = - f(r)\ud t^2&  + [f(r)]^{-1} \ud r^2 +r^2(\ud \theta^2 +\sin^2(\theta) \ud \phi^2),\\\label{SdS}
& f(r)\equiv1-\frac{2m}{r}- \frac{\Lambda r^2}{3},
\end{align}
which represents a pointlike gravitational lens of mass $m$ embedded within a de Sitter cosmology.\footnote{Being a static, spherically symmetric solution of the cosmological vacuum Einstein field equations ($G_{\mu\nu}+\Lambda g_{\mu\nu}=0$) the SdS metric (\ref{SdS}) clearly generalises the Schwarzschild black hole to account for the presence of $\Lambda$. That said, without an asymptotically flat spatial limit $r\to \infty$ at which to evaluate the ADM mass \cite{ADM59}, it is unclear whether $m$ remains the mass of the system when $\Lambda\ne0$. Indeed, in the absence of the ADM definition, a reasonable approach might be to define the mass of an uncharged static black hole by the area $A$ of its event horizon: $m_\mathrm{actual}\equiv \sqrt{A/16\pi}=m(1+ 4\Lambda m^2/3+ \ldots)$. Fortunately, this correction is negligible even for very massive objects: e.g.\ galaxy clusters have $\Lambda m^2\sim 10^{-16}$. Moreover, this argument applies to any analytic definition $m_\mathrm{actual}\equiv m_\mathrm{actual}(m,\Lambda)$: provided $m_\mathrm{actual}(m,0)=m$, dimensional considerations guarantee $m_\mathrm{actual}=m(1+ O(\Lambda m^2))$. Hence $m$ can be thought of as the true mass for all practical considerations.} In order to determine the paths taken by light in this spacetime, we restrict our interest to the ``equatorial plane'' $\theta=\pi/2$ and use standard techniques (e.g.\ \cite{Hobson06}) to obtain the orbital equation for null geodesics:
\begin{align}\label{orbit}
\left(\frac{\ud^2 }{\ud \phi^2} + 1\right)\frac{1}{r} &= \frac{3m}{r^2}.
\end{align}
Note that the conspicuous absence of $\Lambda$ from this equation is not the result of an approximation: the cosmological constant has no effect on the motion of light in the $(r,\phi)$ coordinate plane. We now seek approximate solutions of the orbital equation (\ref{orbit}) which approach the lens, reach a minimum radius $r_\mathrm{min}$ (by convention, at $\phi=\pi/2$), and then move away. Because (\ref{orbit}) is invariant under $\ud \phi \to -\ud \phi$, such paths must be symmetric about the point of closest approach $\phi=\pi/2$. It is a simple task to verify that
\begin{align}\label{lightpath}\nonumber
\frac{R}{r(\phi)} &= \sin (\phi) + \frac{3m}{2R} \left( 1 + \frac{ \cos (2\phi)}{3}\right)\\\nonumber
&\quad {} - \frac{3 m^2}{4R^2}\left( 5(\phi-\pi/2) \cos(\phi) + \frac{1}{4}\sin (3\phi)\right)\\
&\quad {}+ O(m^3/R^3),
\end{align}
are the solutions we seek: these paths display the desired symmetry about $\phi=\pi/2$ and solve the orbital equation (\ref{orbit}) to the stated accuracy. The paths are parametrised by a constant $R$ that roughly corresponds to the distance of closest approach: $r_\mathrm{min}= r(\pi/2)=R(1+O(m/R))$. We assume $R\gg m$; this justifies the expansion of (\ref{lightpath}) in powers of $m/R$ and ensures that the light path avoids the event horizon near $r= 2m$. Equation (\ref{lightpath}) improves upon the accuracy of the solution used by Rindler and Ishak \cite{Rindler07} by including terms of order $m^2/R^2$; these higher-order corrections will not be crucial to our argument, but they will allow us to extend our conclusions beyond the limits of the standard approximation.

Before exploring the observational implications of (\ref{lightpath}) in the presence of $\Lambda$, it will be useful to refamiliarise ourselves with the approach taken when $\Lambda=0$.

\begin{figure}
\includegraphics[scale=.45]{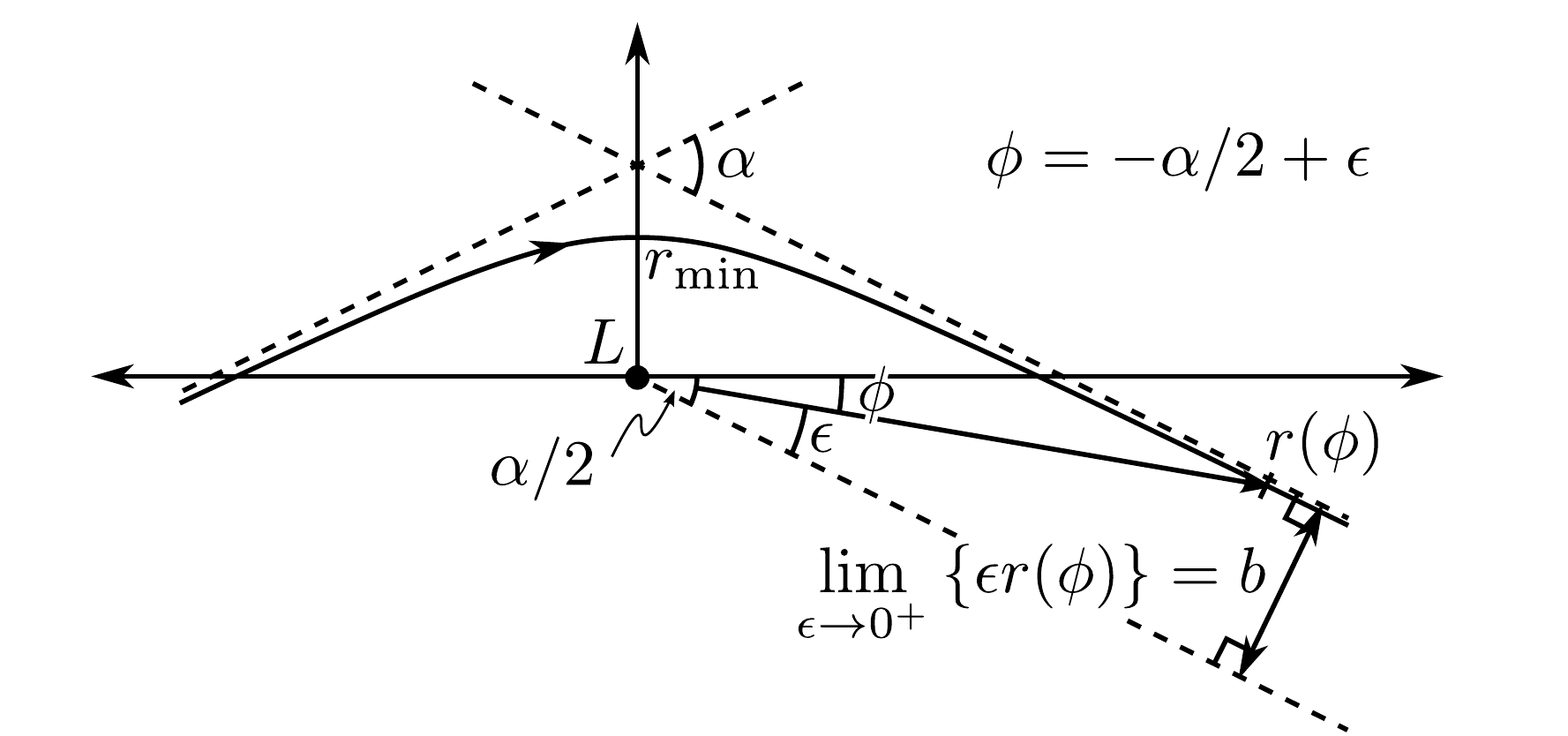}
\caption{When $\Lambda =0$, the bending angle $\alpha$ and impact parameter $b$ are defined in the asymptotically flat spatial limit $r\to\infty$. The symmetry of the light path (\ref{lightpath}) about $\phi=\pi/2$ implies that spatial infinity is approached as $\phi\to -\alpha/2$ from above, where the asymptotic behaviour is $r(\phi)=r(-\alpha/2+\epsilon)\sim b/\epsilon$.}\label{asymp}
\centering
\end{figure}

\section{\texorpdfstring{$\Lambda =0$}{Lambda=0}}\label{nolambda}
When there is no cosmological constant, the SdS metric (\ref{SdS}) is asymptotically flat as $r\to \infty$. This is allows us to treat the lensing process as a ``black box'' that takes an incoming beam of light, with some impact parameter $b>0$, and produces an outgoing beam of light deflected by an angle $\alpha>0$. As figure \ref{asymp} indicates, the values of $\alpha$ and $b$ are determined by the asymptotic behaviour of the light path: $r(-\alpha/2 +\epsilon)\sim b/\epsilon$ as $\epsilon \to 0^+$. To apply this limit to the path (\ref{lightpath}), first note that $r(-\alpha/2)=\infty$ implies $\alpha\sim O(m/R)$;  hence equation (\ref{lightpath}) becomes
\begin{align}\nonumber
\frac{R\epsilon}{b}&= -\frac{\alpha}{2} + \frac{2m}{R} + \frac{15 \pi m^2 }{8R^2} + \epsilon\left(1+O(m^2/R^2) \right)\\
&\quad {} +O(\epsilon^2) + O(m^3/R^3),
\end{align}
for $\epsilon>0$. Solving the part of this equation that is independent of $\epsilon$, and then the part linear in $\epsilon$, we obtain
\begin{align}\bs
\alpha &= \frac{4m}{R}+ \frac{15\pi m^2}{4R^2}   + O(m^3/R^3),\\
R &= b \left(1+ O(m^2/R^2) \right),\es
\end{align}
and thus
\begin{align}\label{oldalpha}
\alpha = \frac{4m}{b}+\frac{15\pi m^2}{4b^2} + O(m^3/b^3).
\end{align}
This formula characterises the black box of gravitational lensing. The first term $4m/b$ is well known, widely used, and sufficiently accurate for most applications; the second term $15\pi m^2/4b^2$ is due to the corrections of order $m^2/R^2$ included in equation (\ref{lightpath}). Although this correction is too small to be observed in most situations, the uncertainties that arise in our treatment of $\Lambda$ will be even smaller, comparable to the terms $O(m^3/b^3)$ that have been discarded.

\begin{figure}
\includegraphics[scale=.45]{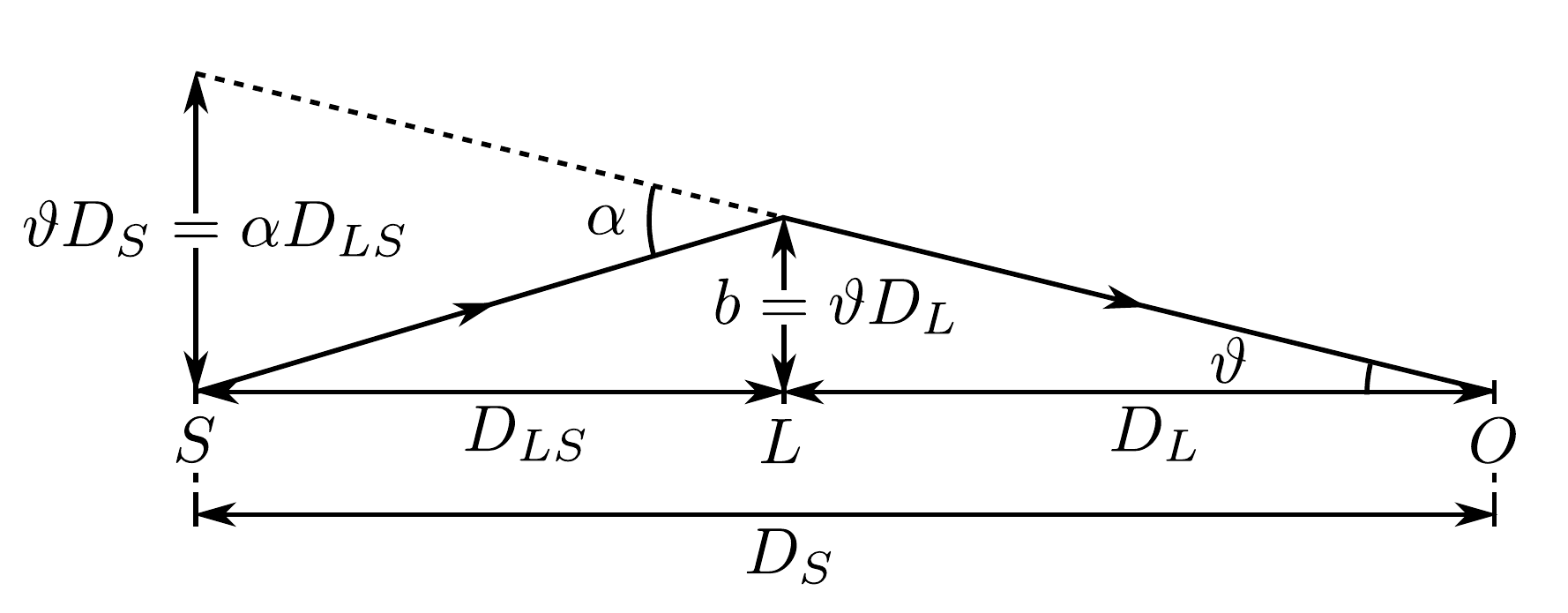}
\caption{When $b\ll D_L,D_{LS}$, gravitational lensing can be treated as an instantaneous event along the light path. For the case where the light source $S$, lens $L$, and observer $O$ are collinear, the geometry of this process is summarised by the figure above. Note that $\{D_L, D_S, D_{LS}\}$ are \emph{unlensed} angular diameter distances, and that the small angle approximation has been applied. This picture is only physically accurate when $\Lambda=0$; nonetheless, it can also be used to \emph{define} effective values of $\alpha$ and $b$ when $\Lambda\ne0$.}\label{triangles}
\centering
\end{figure}

To apply (\ref{oldalpha}) to astrophysical observations, one recognises that light is only appreciably affected by the lens within a region $r\approx r_\mathrm{min} \approx b$ that is much smaller than the distances between the light source $S$, the lens $L$, and the observer $O$. This allows us to treat the whole lensing process as an instantaneous ``event'' along the light path: the lensing law (\ref{oldalpha}) is applied at that single point, and the effect of the lens is ignored everywhere else. When $S$, $L$ and $O$ are collinear, this scheme is represented by figure \ref{triangles}; note that the small angle approximation (justified by $b\ll D_L, D_{LS}$) neglects fractional errors of order $O(\vartheta^2)$ and $O(\alpha^2)$ and so operates the same level of accuracy as the lensing law (\ref{oldalpha}). A key feature of this model is that, because we have restricted the effect of the lens to the single deflection point, the angular diameter distances $\{D_L, D_S, D_{LS}\}$ take on the ``unlensed'' values that would be measured if the lens mass were zero. (We will return to the precise meaning of ``unlensed'' in section \ref{unlensed} and consider ``lensed'' distances in the appendix.)

In this fashion, the lensing law (\ref{oldalpha}) and figure \ref{triangles} constitute the classical model of gravitational lensing; together, they relate the observable $\vartheta$ to the physical quantities of interest: $\{D_L, D_S, D_{LS}\}$ and $m$. Even though the derivation of (\ref{oldalpha}) clearly required the lens to be embedded in an asymptotically flat spacetime, this model is often na\"ively applied to cosmological settings when this fundamental assumption no longer applies. To determine whether this na\"ivety is actually a problem, we must examine how the above model is altered by the presence of a positive cosmological constant.

\section{\texorpdfstring{$\Lambda >0$}{Lambda>0}}\label{yeslambda}
Although the cosmological constant does not appear in the orbital equation (\ref{orbit}) or the light path (\ref{lightpath}), the presence of $\Lambda$ alters the analysis of gravitational lensing in four key ways:
\begin{enumerate}
\item With $\Lambda >0$, the SdS metric (\ref{SdS}) is no longer asymptotically flat as $r\to\infty$; indeed, we encounter a cosmological horizon near $r=\sqrt{3/\Lambda}$. Consequently, we can no longer treat the lensing processes as a black box that takes an incoming beam from infinity and produces an outgoing beam at infinity. We cannot even \emph{define} the bending angle $\alpha$ and impact parameter $b$ in the usual way, because there is no flat spatial limit at which to measure them. (This is \emph{not} a coordinate artefact: it is a feature of the de Sitter geometry. Parallel beams of light  \emph{diverge} as they move away from the lens, precluding the construction of $\alpha$ and $b$ in this limit.)
\item Physical angles are computed using the SdS metric (\ref{SdS}), which depends on $\Lambda$.
\item The lens is embedded within a de Sitter cosmology, the comoving observers of which are receding from the lens. This motion leads to relativistic aberration, modifying the angles that are actually observed.
\item The conversion between coordinates and angular diameter distances $\{D_L, D_S, D_{LS}\}$ will also depend on $\Lambda$. Furthermore, there is an unavoidable ambiguity in assigning these distances \emph{unlensed} values.
\end{enumerate}
The first two of these considerations were championed by Rindler and Ishak, serving as the main impetus for their original paper \cite{Rindler07}. However, their analysis was incomplete: they failed to consider the last two aspects of the problem, as listed above. Indeed, one finds these aspects have generally been marginalised or ignored in the literature; for instance, Sch\"ucker ignores cosmological aberration \cite{Schucker09a} and Sereno fails to account for the ambiguity of unlensed distances \cite{Sereno09}. We will treat all four aspects in turn and then make a full comparison with the $\Lambda=0$ case.

\subsection{No Asymptotic Flatness}\label{noAF}
\begin{figure}
\includegraphics[scale=.45]{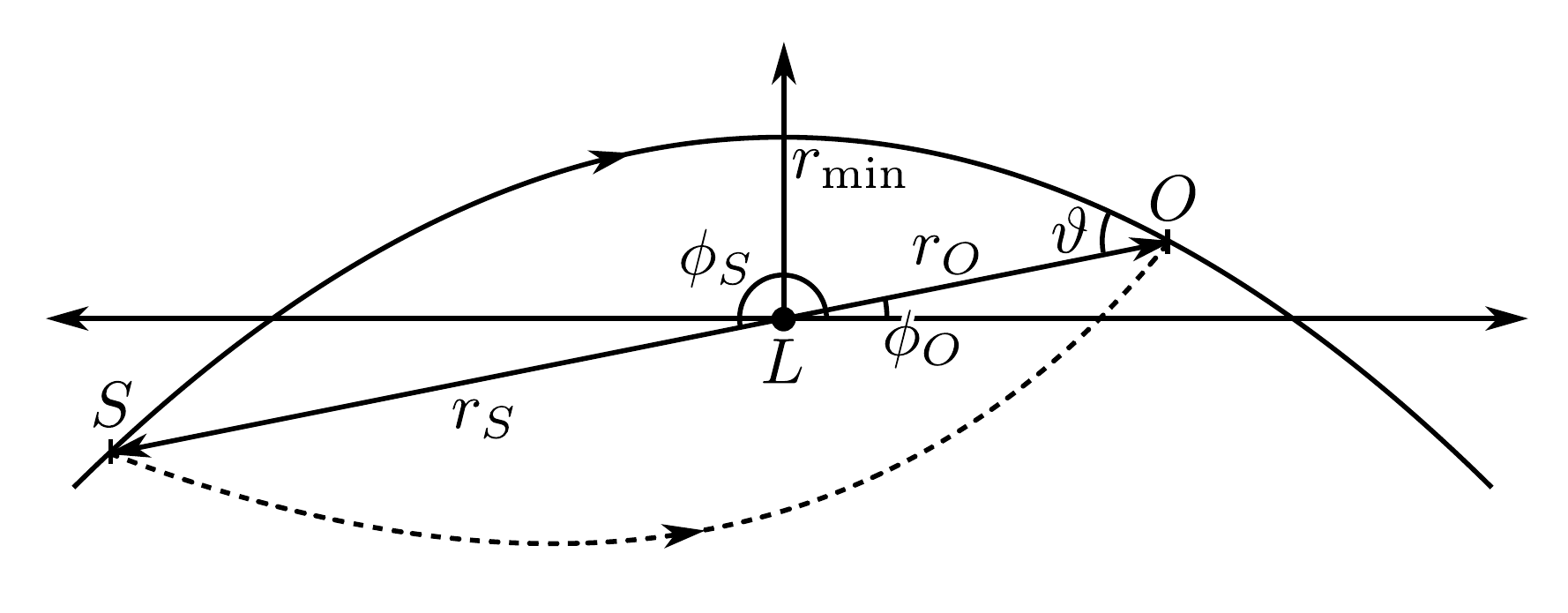}
\caption{Schematic of the path (\ref{lightpath}) of light emitted by a source $S$, lensed by $L$, and observed at $O$, when these three points are collinear.  Due to the axis of symmetry $SO$, light will take many similar paths from $S$ to $O$ (one of which is indicated with a dotted line) and the image of $S$ at $O$ will be an Einstein ring of angular radius $\vartheta$. Note that the proportions of this diagram have been distorted in the interest of readability: in actuality, $r_O,r_S\gg r_\mathrm{min}$ and $\vartheta,\phi_O \ll 1$.}\label{finitedist}
\centering
\end{figure}

In the absence of asymptotic flatness and the black box approach it justifies, we are forced to model the lensing process more realistically: light will arrive from the source $S$ at some \emph{finite} radius $r_S<\sqrt{3/\Lambda}$ and meet the observer $O$ at some \emph{finite} radius $r_O<\sqrt{3/\Lambda}$. This scheme is illustrated in figure \ref{finitedist}, where we have once again focused on the case where  $S$, $L$, and $O$ are collinear. This more accurate picture replaces both figures \ref{asymp} and \ref{triangles} from the $\Lambda=0$ analysis.

In order to make predictions from this model, our first step will be to obtain a formula for the quantity
\begin{align}\label{r'}
r'_O\equiv \left.\frac{\ud r(\phi)}{\ud\phi}\right|_{\phi=\phi_O},
\end{align}
which is closely related to the observable $\vartheta$. To evaluate (\ref{r'}) we first notice that the symmetry of the light path (\ref{lightpath}) about $\phi=\pi/2$ implies
\begin{align}\nonumber
r_S\equiv r(\phi_S)= r(\pi +\phi_O) &= r(\pi/2+(\pi/2 +\phi_O))\\\nonumber
&=r(\pi/2-(\pi/2 +\phi_O))\\
&=r(-\phi_O),
\end{align}
and so it follows from (\ref{lightpath}) that
\begin{align}\label{r-}
\frac{R}{r_O}-\frac{R}{r_S}&= 2\sin (\phi_O)+ O(m^2/R^2),
 \\\nonumber
\frac{R}{r_O}+\frac{R}{r_S}&= \frac{3 m}{R}\left(1 + \frac{\cos (2\phi_O)}{3}\right) \\\label{r+}
&\quad {} + \frac{15\pi m^2}{4R^2} \cos( \phi_O)  + O(m^3/R^3).
\end{align}
Note that (\ref{r+}) implies $R^2= O(mr)$ and therefore (\ref{r-}) gives $\sin(\phi_O)= O(R/r)=O(\sqrt{m/r})$. With these magnitudes in mind, we can rewrite equation (\ref{r+}) as
\begin{align}\nonumber
\frac{R}{r_O}+\frac{R}{r_S}&= \frac{3 m}{R}\left(1 + \frac{1-2\sin^2(\phi_O)}{3}\right) \\\nonumber
&\quad {} + \frac{15\pi m^2}{4R^2} \sqrt{1- \sin^2(\phi_O)}  + O(m^3/R^3)\\\label{cubicR}
&= \frac{4m}{R} + \frac{15\pi m^2}{4R^2} + O(m^{3/2}/r^{3/2}).
\end{align}
This is a cubic equation for $R$, which is solved to sufficient accuracy by 
\begin{align}\label{Rsol}
R=\frac{2m^{1/2}}{\sqrt{r_O^{-1}+ r_S^{-1}}} + \frac{15\pi m}{32} + O(m^{3/2}/r^{1/2}),
\end{align}
as can be checked by direct substitution.

We now differentiate (\ref{lightpath}) with respect to $\phi$ and evaluate the resulting equation at $\phi=\phi_O$:
\begin{align}\nonumber
-\frac{R r'_O}{r^2_O}&= \cos(\phi_O) - \frac{m}{R}\sin(2\phi_O) + O(m^2/R^2)\\
&=1 + O(m/r).
\end{align}
Combining this with equation (\ref{Rsol}) we have
\begin{align}\nonumber
-\frac{ r_O}{r'_O}&= \frac{R}{r_O}\left(1 + O(m/r)\right)\\\label{r/r'}
&= \frac{2m^{1/2}}{r_O\sqrt{r_O^{-1}+ r_S^{-1}}} + \frac{15\pi m}{32 r_O} + O(m^{3/2}/r^{3/2}).
\end{align}
This is the formula for $r'_O$ that we need; our next task is to relate the left-hand side of (\ref{r/r'}) to $\vartheta$, the angular radius of the Einstein ring observed at $O$.

\subsection{Physical Angles}\label{Physangles}
\begin{figure}
\includegraphics[scale=.45]{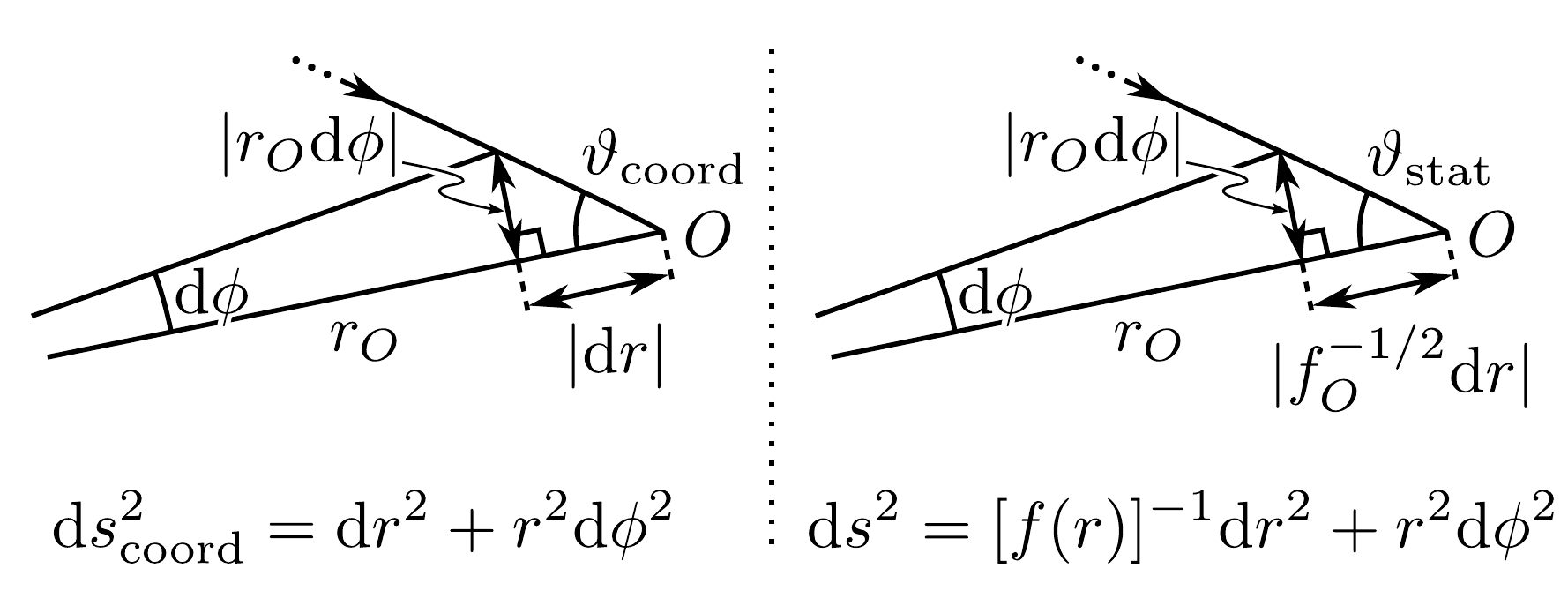}
\caption{Two views of the angle $\vartheta$ formed at $O$ in figure \ref{finitedist}. On the left, the geometry has been projected onto the flat $(r,\phi)$ coordinate plane; on the right, it takes place within the physical space formed by restricting the SdS metric (\ref{SdS}) to $t=\mathrm{const}$, $\theta= \pi/2$. As the light path (\ref{lightpath}) forms the hypotenuse of the infinitesimal triangle,  $\ud r= r'_O \ud \phi$.}\label{angles}
\centering
\end{figure}

It is important to distinguish between three different definitions of angles in the Schwarzschild--de Sitter spacetime, and hence view the angle $\vartheta$ in figure \ref{finitedist} as a shorthand for three different quantities: $\{\vartheta_\mathrm{coord}, \vartheta_\mathrm{stat},\vartheta_\mathrm{obs}\}$. $\vartheta_\mathrm{coord}$ is the angle formed in the $(r,\phi)$ coordinate plane, with a \emph{flat} metric $\ud s^2_\mathrm{coord}=\ud r^2 + r^2 \ud \phi^2$. This angle is a convenient mathematical construction, but has no obvious physical meaning. In contrast, $\vartheta_\mathrm{stat}$ represents the \emph{physical} angle formed in the curved space that results from fixing $t=\mathrm{const}$, $\theta= \pi/2$ in the SdS metric (\ref{SdS}): $\ud s^2= [f(r)]^{-1} \ud r^2 + r^2 \ud \phi^2$. The difference between these definitions is depicted in figure \ref{angles}, from which it also follows that
\begin{align}\bs\label{thetas}
\tan (\vartheta_\mathrm{coord})&=\frac{r_O}{|r'_O|},\\
\tan (\vartheta_\mathrm{stat})&=\frac{r_O }{|r'_O|}f_O^{1/2},
\es
\end{align}
where $f_O\equiv f(r_O)$.  Because $\vartheta_\mathrm{stat}$ has been constructed on the spatial hypersurface $t=\mathrm{const}$, it represents the angle that would be measured by an observer who is \emph{stationary} with respect to the coordinates $(t,r,\theta,\phi)$ used to represent the SdS metric (\ref{SdS}). In other words, $\vartheta_\mathrm{stat}$ is the angle as seen by an observer with 4-velocity 
\begin{align}
u_\mathrm{stat}^{\mu}=(f_O^{-1/2},0,0,0).
\end{align}
Of course, we are not particularly interested in observers that are stationary in the coordinates $(t,r,\theta,\phi)$: we actually wish to know $\vartheta_\mathrm{obs}$, the angle as measured by a \emph{cosmological observer} at $O$, with 4-velocity $u^\mu_\mathrm{obs}$. 

We will determine $u^\mu_\mathrm{obs}$ in the next section. At present it suffices to say that  cosmological observers must respect the spherical symmetry of the Schwarzschild--de Sitter spacetime (\ref{SdS}) and so can only move \emph{radially}. Hence $\vartheta_\mathrm{obs}$ will be related to $\vartheta_\mathrm{stat}$ by the standard relativistic aberration formula: 
\begin{align}\label{abberation}
\cos(\vartheta_\mathrm{obs})=\frac{\cos(\vartheta_\mathrm{stat}) -v }{1- v \cos(\vartheta_\mathrm{stat})},
\end{align}
where $v$ is the \emph{outward} radial 3-velocity of $u^\mu_\mathrm{obs}$, relative to $u_\mathrm{stat}^\mu$. Note that (\ref{abberation}) implies
\begin{align}
\tan(\vartheta_\mathrm{obs})= \frac{\sin(\vartheta_\mathrm{stat})\sqrt{1-v^2}}{\cos(\vartheta_\mathrm{stat})- v },
\end{align}
which becomes
\begin{align}
\vartheta_\mathrm{obs} = \sqrt{\frac{1+v}{1-v}}\vartheta_\mathrm{stat} + O(\vartheta^3_\mathrm{obs}),
\end{align}
in the small angle approximation. Using (\ref{thetas}) we therefore arrive at
\begin{align}\label{theta1}
\vartheta_\mathrm{obs} = \sqrt{\frac{1+v}{1-v}} \frac{r_O}{|r_O'|}f^{1/2}_O+ O(\vartheta^3_\mathrm{obs}),
\end{align}
which achieves our aim of relating $\vartheta_\mathrm{obs}$ to the left-hand side of equation (\ref{r/r'}). In the next section, we will determine $u^\mu_\mathrm{obs}$ and hence $v$.

\subsection{Cosmological Observers}\label{observers}
A Friedmann-Robertson-Walker (FRW) cosmology is foliated by spatial hypersurfaces over which the universe is homogeneous and isotropic; ``Copernican'' cosmological observers then follow the timelike geodesics normal to these hypersurfaces. Of course, in the Schwarzschild--de Sitter spacetime (\ref{SdS}) the central mass inevitably breaks spatial homogeneity, so there can be no \emph{truly} Copernican observers. Nonetheless, there is a natural generalisation of this idea that remains applicable: the timelike geodesic congruence that (i) respects all the continuous symmetries of the spacetime, and (ii) approaches \emph{local} isotropy as rapidly as possible as $r$ becomes large. We take this congruence, being the ``most Copernican'' available, as the cosmological observers (with 4-velocity $u_\mathrm{obs}^\mu$ at $O$) that define $\vartheta_\mathrm{obs}$.

To calculate $u_\mathrm{obs}^\mu$, we begin with the most general vector field that respects the spherical symmetry and time-invariance of the metric (\ref{SdS}): $u^\mu=(a(r), b(r), 0, 0)$. This field will be the unit tangent vector of the geodesic congruence we seek; in other words, freely falling cosmological observers will move along paths $x^\mu(\tau)$ given by $\ud x^{\mu}/\ud \tau=u^\mu$. Writing $a(r)= E(r)/ f(r)$, we enforce the timelike normalisation condition $u^\alpha u_\alpha=-1$, followed by the geodesic equation $ u^\alpha \nabla_\alpha u^\mu=0$, and arrive at
\begin{align}\label{u}
u^\mu = (E/f(r),\pm \sqrt{E^2 - f(r)},0,0),
\end{align}
where $E$ is a constant. Clearly we must take $E>0$ for $u^\mu$ to be future directed, and must also discard the negative root, as it describes a contracting cosmology. Although $u^t$ diverges where $f(r)=0$ (i.e.\ at the black-hole and cosmological horizons) this is simply a coordinate artefact: coordinate-independent quantities derived from $u^\mu$ will remain finite at the horizons. In contrast, imaginary values of $u^r$ are clearly \emph{not} coordinate artefacts, so if the congruence is to exist for all positive values of $r$ (or at least, for all values of $r$ between the black-hole and cosmological horizons) we must insist that $E^2\ge E_\mathrm{min}^2 \equiv \max_{r>0} \{f(r)\}= 1 - (9m^2 \Lambda)^{1/3}$.

Finally, to quantify the local anisotropy of the congruence, we evaluate the shear tensor
\begin{align}
\sigma^{\mu}{}_{\nu}\equiv \frac{1}{2} (\nabla^{\mu}u_{\nu} +\nabla_{\nu}u^{\mu})  - \frac{1}{3}( \nabla_{\alpha}u^{\alpha})(\delta^\mu_{\nu} + u^\mu u_\nu),
\end{align}
which measures the anisotropic deformation of a sphere of geodesics near the observer (see  \cite{Wald}). The local anisotropy of the congruence is then fully characterised by the eigenvalues of $\sigma^{\mu}{}_{\nu}$: $\{0,-2\sigma_*,\sigma_*,\sigma_*\}$, where\footnote{The pattern of eigenvalues can be inferred from (i) $ \sigma^{\mu}{}_{\alpha}u^\alpha=0$, (ii) $\sigma^{\alpha}{}_{\alpha}=0$, and (iii) the spherical symmetry of the system. The easiest way to determine the magnitude of $\sigma_*$ is to evaluate $\sigma^2\equiv\sigma^{\alpha}{}_{\beta}\sigma^{\beta}{}_{\alpha}$ using Raychaudhuri's equation \cite{Ray55}, noting that the twist tensor vanishes: $\omega_{\mu\nu}\equiv (\nabla_{\nu}u_{\mu} -\nabla_{\mu}u_{\nu})/2=0$.}
\begin{align}
\sigma_* = \frac{3m + r(E^2-1)}{3r^2\sqrt{E^2 -f(r)}}.
\end{align}
With $E^2\ge E^2_\mathrm{min}$, this coordinate independent quantity will not diverge at any value of $r>0$. Hence we can meaningfully continue past the cosmological horizon and consider the asymptotic behaviour of $\sigma_*$ as $r\to \infty$:
\begin{align}\nonumber
\sigma_* &\sim 
\frac{(E^2-1)}{ r^2\sqrt{3\Lambda}}&\mbox{if } E^2 &\ne 1,\\
\sigma_*&\sim\frac{m\sqrt{3/\Lambda}}{r^{3}} &\mbox{if } E^2 &= 1.
\end{align}
We conclude that $E^2=1$ specifies the congruence that approaches local isotropy most rapidly and hence 
\begin{align}\label{u1}
u^\mu = ([f(r)]^{-1}, \sqrt{1 - f(r)},0,0)
\end{align}
describes the congruence of cosmological observers we seek. At $O$, the observer's 4-velocity is therefore
\begin{align}\label{uobs}
u_\mathrm{obs}^\mu = (f^{-1}_O, \sqrt{1 - f_O},0,0),
\end{align}
and the value of $v$ (the outward radial 3-velocity, relative to $u^\mu_\mathrm{stat}$) can be obtained via the Lorentz factor: 
\begin{align}
- u^\mu_\mathrm{obs}u^\nu_\mathrm{stat}g_{\mu\nu}= \gamma = (1-v^2)^{-1/2},
\end{align}
which yields
\begin{align}\label{v}
v=\sqrt{1-f_O}.
\end{align}
Note that $v|_{m=0}= r_O \sqrt{\Lambda/3}$ describes the truly Copernican observers of de Sitter spacetime, as one would expect. (Of course, \emph{actual} gravitational lenses are embedded within a $\Lambda$CDM cosmology, the observers of which differ from those derived here. Such complications are besides the point, however: SdS is the extreme case where $\Lambda$ dominates, and thus the effect of $\Lambda$ on lensing should be strongest -- we use it as is a mathematical \emph{litmus test} for the existence of a $\Lambda$--lensing effect in general, not as a physically realistic model.)

\subsection{Unlensed Distances}\label{unlensed}
The final ingredients needed for our analysis are the unlensed angular diameter distances $\{D_L, D_S, D_{LS}\}$. (Recall that we use \emph{unlensed} values so a direct comparison can be made with the $\Lambda=0$ analysis of section \ref{nolambda}. We will explore the use of \emph{lensed} distances in the appendix, and find that our conclusions are only strengthened by this modification.) Clearly one cannot define unlensed distances without mapping points in the actual spacetime (\ref{SdS}) onto points in a hypothetical ``lens-free'' spacetime [de Sitter space: (\ref{SdS}) with $m=0$]. Because these two spaces have different geometries, the details of this map will be \emph{arbitrary} to some extent: there will be a choice over which geometrical relationships the map will preserve. For the case at hand, both spaces are spherically symmetric and time-invariant, and our map must respect this. Consequently, our freedom is limited to the function that maps the $r$ coordinate of the actual $m\ne0$ space onto the $r$ coordinate of the hypothetical $m=0$ space: $r \mapsto r_{m=0}$. In selecting this function there is an unavoidable tension between (i) preserving the area of spheres centred on $L$, and (ii) preserving the velocities and redshifts of cosmological observers (\ref{u1}). If we wish to preserve areas, we must use the identity map $r_{m=0}=r$; if we wish to preserve cosmological velocities and redshifts, we must use the map such that $f_{m=0}(r_{m=0})=f(r)$. Without a fundamental reason to adopt one convention over the other, we are forced to accept that both options (and all possibilities in between) are equally valid. 
Hence there is an unavoidable \emph{geometric ambiguity} in the locations of $S$ and $O$ in the lens-free space, and a corresponding uncertainty in the unlensed values of $\{D_L, D_S, D_{LS}\}$.

\begin{figure}
\includegraphics[scale=.45]{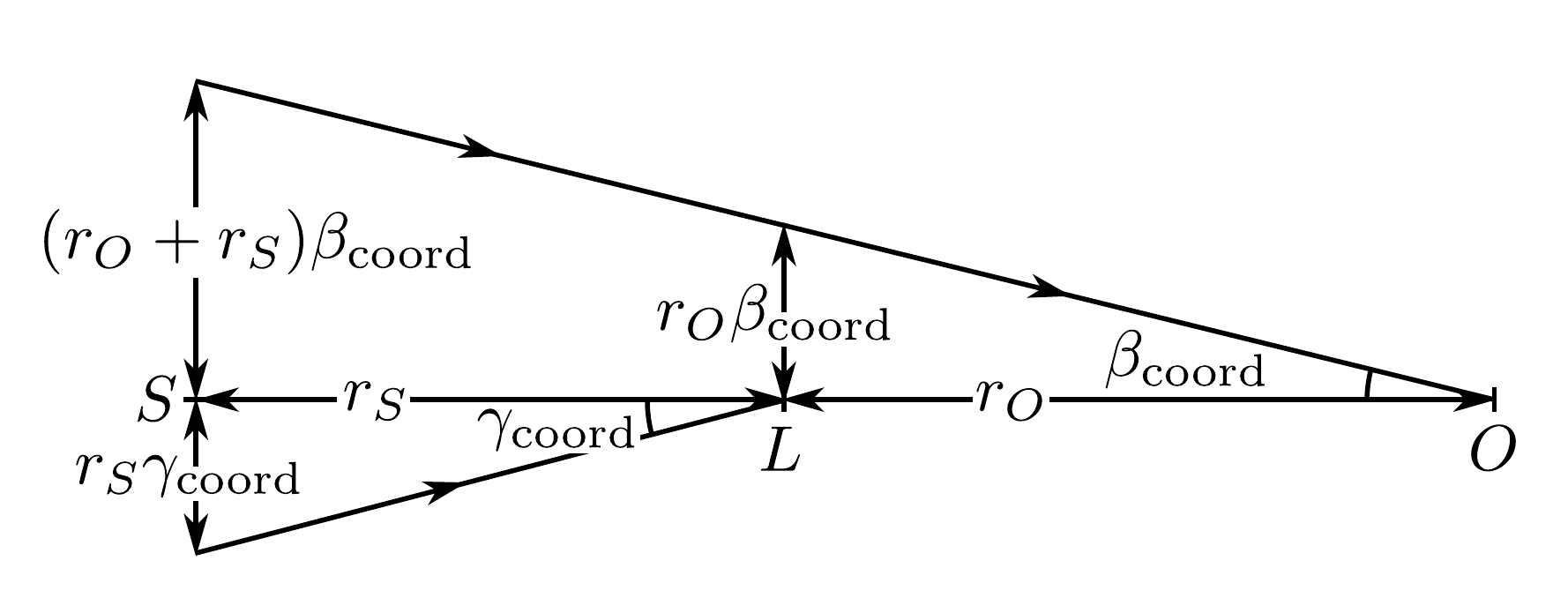}
\caption{When $m=0$, light follows straight lines in the $(r,\phi)$ coordinate plane; hence the above triangles determine the unlensed angular diameter distances $\{D_L, D_S,D_{LS}\}$ in the small angle limit $\beta,\gamma \to0$. Note that the physical heights of the triangles are equal to their heights in the coordinate plane: this is because the heights at $S$ are purely tangential (so $\ud s=|r \ud \phi|=\ud s_\mathrm{coord}$) and because $f_{m=0}=1$ at $L$. 
}\label{Dtriangles}
\centering
\end{figure}

Setting this issue aside for the moment, let us provisionally adopt the area-preserving  convention and determine the unlensed distances that it assigns. Because light travels along straight lines in the $(r,\phi)$ coordinate plane of the lens-free space (as follows from the orbital equation (\ref{orbit}) with $m=0$) the values of $\{D_L, D_S,D_{LS}\}$ can be inferred from the right-angled triangles depicted in figure \ref{Dtriangles}. Accounting for the aberration factors discussed in section \ref{Physangles}, and making use of (\ref{v}), we have
\begin{align}\nonumber
\beta_\mathrm{obs}&= \beta_\mathrm{coord} \left[\sqrt{\frac{1+v}{1-v}}f^{1/2}_O\right]_{m=0}\\
&= \beta_\mathrm{coord}(1+ r_O \sqrt{\Lambda/3}),
\end{align}
and, trivially, $\gamma_\mathrm{obs}=\gamma_\mathrm{coord}$. Hence
\begin{align}\label{Ds}\bs
D_L &= \frac{ r_O\beta_\mathrm{coord} }{\beta_\mathrm{obs}}= \frac{r_O}{1+r_O\sqrt{\Lambda/3}},\\
D_S &= \frac{(r_O +r_S)\beta_\mathrm{coord} }{\beta_\mathrm{obs}}= \frac{r_O +r_S}{1+r_O\sqrt{\Lambda/3}},\\
D_{LS} &=\frac{r_S \gamma_\mathrm{coord}}{\gamma_\mathrm{obs}} =r_S\es
\end{align}
are the unlensed angular diameter distances ascribed by the area-preserving map $r_{m=0}=r$.

Had we adopted the redshift-preserving convention, $S$ and $O$ would be placed at slightly different radii in the lens-free space, altering the values of $\{D_L, D_S,D_{LS}\}$. In general, moving between the two conventions introduces a radial shift $\delta r\equiv r_\mathrm{(2)}- r_\mathrm{(\mathrm{1})}$, where
\begin{align}
 r_\mathrm{(\mathrm{1})}&=r,& f_{m=0}(r_\mathrm{(2)})&= f(r).
\end{align}
This implies
\begin{align}
\delta r =   \frac{3m}{\Lambda r^2},
\end{align}
working to first order in $\delta r$. Therefore we may use the angular diameter distances (\ref{Ds}) provided we understand them as having a fractional uncertainty
\begin{align}\label{geoamb}
\delta D/D= O(\delta r/r)=O(m/\Lambda r^3),
\end{align}
arising from the ambiguity involved in locating $S$ and $O$ in the lens-free de Sitter space. We note that this uncertainly has been broadly ignored in the literature, even when the effects of interest are smaller than the geometric ambiguity \cite{Sereno09}.

\subsection {Synthesis}\label{Synth}
Let us now assemble the various elements obtained in the previous sections; from these we will construct a model of gravitational lensing, valid for the $\Lambda>0$ Schwarzschild--de Sitter spacetime (\ref{SdS}), that will replace the na\"ive $\Lambda=0$ analysis of section \ref{nolambda}.

We begin by inserting (\ref{r/r'}) and (\ref{v}) into (\ref{theta1}):
\begin{align}\label{thetaobs}\nonumber
\vartheta_\mathrm{obs} &= \left(1+\sqrt{1-f_O}\right)\left( \frac{2m^{1/2}}{r_O\sqrt{r_O^{-1}+ r_S^{-1}}} + \frac{15\pi m}{32 r_O} \right)\\
&\quad{}+ O(\vartheta^3_\mathrm{obs}).
\end{align}
Approximating $\sqrt{1-f_O}=r_O\sqrt{\Lambda/3}+O(m/\Lambda^{1/2}r^2)$, equation (\ref{thetaobs}) becomes
\begin{align}\label{thetaobs2}\nonumber
\vartheta_\mathrm{obs} &= \frac{1+r_O\sqrt{\Lambda/3}}{r_O}\left( \frac{2m^{1/2}}{\sqrt{r_O^{-1}+ r_S^{-1}}} + \frac{15\pi m}{32} \right)\\
&\quad{}+ O(\vartheta^3_\mathrm{obs}/\Lambda^{1/2}r),
\end{align}
where $\vartheta_\mathrm{obs}\sim O(\sqrt{m/r})$ and $r_O,r_S<\sqrt{3/\Lambda}$ were used to simplify the error term. Expressing the coordinates $\{r_O,r_S\}$ in terms of unlensed angular diameter distances (\ref{Ds}), we arrive at
\begin{align}\label{thetaobsD}
\vartheta_\mathrm{obs} = 2 \sqrt{\frac{m D_{LS}}{D_L D_S}} + \frac{15\pi m}{32 D_L} + O(\vartheta^3_\mathrm{obs}/\Lambda r^2),
\end{align}
where the dominant error term stems from the unavoidable ambiguity (\ref{geoamb}) of unlensed distances. Thus, working to the maximum accuracy that unlensed distances allow, we see that $\vartheta_\mathrm{obs}$ can be expressed as a function of $m$ and $\{D_L, D_S,D_{LS}\}$ alone, without any dependence on $\Lambda$. This strongly supports the traditional view: the observable effects of $\Lambda$ are \emph{already accounted for} within the angular diameter distances. 

It is tempting to end our analysis at this point, taking equation (\ref{thetaobsD}) as proof that ``Lambda does not lens''. However, to stop at this stage would only inspire the following question: why focus on equation (\ref{thetaobsD}) and not equation (\ref{thetaobs2})? We cannot dismiss equation (\ref{thetaobs2}) as somehow ``less physical'' than (\ref{thetaobsD}): the coordinates $\{r_O,r_S\}$ are also physical distances, specifying the areas of static spheres centred on $L$. As such, it seems that nothing prevents us from halting our analysis at (\ref{thetaobs2}) and arriving at the opposite conclusion, that $\vartheta_\mathrm{obs}$ really \emph{is} a function of $\Lambda$. It goes without saying that our reasoning should not depend on the particular choice of variables used to express $\vartheta_\mathrm{obs}$. Rather, what we actually need to determine is the effect of $\Lambda$ on the \emph{standard lensing formalism} of section \ref{nolambda}. Consequently, we shall move beyond (\ref{thetaobsD}) and make a direct comparison with the $\Lambda=0$ analysis. To this end, we  run the logic of section \ref{nolambda} in reverse.

Let us take figure \ref{triangles} as our starting point, and use this to \emph{define} effective values of the bending angle and impact parameter:
\begin{align}\bs\label{effdef}
\alpha_{\mathrm{eff}}&\equiv \frac{\vartheta_{\mathrm{obs}} D_S}{D_{LS}} + O(\vartheta_\mathrm{obs}^3),\\
b_{\mathrm{eff}}&\equiv \vartheta_\mathrm{obs} D_L + O(\vartheta_\mathrm{obs}^3 D),\es
\end{align}
the uncertainties reflecting the use of the small angle approximation. Note that, in stark contrast to the bending angle $\alpha$ and impact parameter $b$ of section \ref{nolambda}, $\alpha_{\mathrm{eff}}$ and $b_{\mathrm{eff}}$ are well defined in the absence of asymptotic flatness. We will then use equation (\ref{thetaobsD}) to derive the relationship between $\alpha_\mathrm{eff}$ and $b_\mathrm{eff}$ that holds in the  SdS spacetime. [Alternatively, one could obtain the same result using equations (\ref{thetaobs2}) and (\ref{Ds}).] This relation will replace formula (\ref{oldalpha}) of the $\Lambda=0$ analysis, generalising the black box model of gravitational lensing to $\Lambda>0$. 

Substituting (\ref{thetaobsD}) into the definitions (\ref{effdef}), we obtain
\begin{align}\label{alphaeff}
\alpha_{\mathrm{eff}}&= 2 \sqrt{\frac{m D_S}{D_L D_{LS}}}+ \frac{15\pi m D_S}{32 D_L D_{LS}} + O(\vartheta^3_\mathrm{obs}/\Lambda r^2),
\\\label{beff}
b_{\mathrm{eff}}&= 2 \sqrt{\frac{m D_L D_{LS}}{D_S}} + \frac{15\pi m }{32} +O(\vartheta^3_\mathrm{obs}D/\Lambda r^2)  .
\end{align}
Finally, we rearrange (\ref{beff}),
\begin{align}\sqrt{\frac{D_L D_{LS}}{D_S}}= \frac{b_\mathrm{eff}}{2 \sqrt{m}}\left(1 -\frac{15\pi m}{32 b_\mathrm{eff}} +O\left(\frac{\vartheta^2_\mathrm{obs}}{\Lambda r^2}\right)\right),
\end{align}
and insert this into equation (\ref{alphaeff}): 
\begin{align}\label{newalpha}
\alpha_\mathrm{eff} = \frac{4m}{b_\mathrm{eff}}+\frac{15\pi m^2}{4b_\mathrm{eff}^2} + O\!\left(\frac{m^3}{b_\mathrm{eff}^3 \Lambda r^2}\right),
\end{align}
where the error term was rewritten according to $\vartheta_\mathrm{obs}\sim O(\alpha_\mathrm{eff})=O(m/b_\mathrm{eff})$. Equation (\ref{newalpha}) is the formula  we hoped to derive, encoding the relationship between the effective bending angle and the effective impact parameter (\ref{effdef}) in the Schwarzschild--de Sitter spacetime (\ref{SdS}). Remarkably, our calculation (which accounted for all the effects of the cosmological constant and made no use of asymptotic flatness) has reproduced \emph{exactly the same} relationship as the $\Lambda=0$ asymptotic formula (\ref{oldalpha}).

\section{Conclusions}
According to conventional wisdom, the deflection of light by an isolated point mass can be understood in terms of two simple ingredients: (i) the asymptotic lensing formula (\ref{oldalpha}) of the Schwarzschild spacetime, and (ii) the long-range optical diagram, figure \ref{triangles}. This ``na\"ive'' approach essentially ignores the cosmological constant $\Lambda$: the lens is treated as though embedded within an asymptotically flat space, and angular diameter distances are assumed to account for all cosmological effects, including relativistic aberration. 

In contrast to this simplistic model, we have undertaken a comprehensive and rigorous analysis of lensing in the Schwarzschild--de Sitter spacetime (\ref{SdS}). Accounting for the numerous ways $\Lambda$ influences the lensing process (section \ref{yeslambda}) we have derived an accurate lensing law (\ref{newalpha}) that relates an effective bending angle $\alpha_\mathrm{eff}$ to an effective impact parameter $b_\mathrm{eff}$. These effective quantities are defined (at finite distance from the lens) by figure \ref{triangles}, generalising $\alpha$ and $b$ beyond asymptotic flatness (\ref{effdef}). Hence our accurate description of lensing can also be expressed in two pieces: (i) the effective lensing law (\ref{newalpha}) and (ii)  figure \ref{triangles}, with $\{\alpha_\mathrm{eff},b_\mathrm{eff}\}$ in place of $\{\alpha,b\}$. As these ingredients are mathematically identical to those of the na\"ive model [to the accuracy allowed by the ambiguity (\ref{geoamb}) of unlensed angular diameter distances] it is now clear that the two approaches make exactly the same physical predictions. We conclude that the na\"ive approach, which ignored $\Lambda$, gives a remarkably accurate description of gravitational lensing in the Schwarzschild--de Sitter spacetime. This vindicates the traditional belief that the cosmological constant does not interfere with the standard gravitational lensing formalism \cite{Islam83, Finelli07, Khrip08, Park08, Simpson10} and contradicts the recent claims to the contrary \cite{Rindler07,Ishak08a,Ishak08b, Sereno08,Sereno09,Schucker09a,Schucker09b}.

The only substantive difference between the effective lensing law (\ref{newalpha}) and the na\"ive law (\ref{oldalpha}) is the error term $O(m^3/b_\mathrm{eff}^3 \Lambda r^2)=O(m^5/\Lambda b_\mathrm{eff}^7)$ generated by the inherent ambiguity of unlensed angular diameter distances (section \ref{unlensed}). In the far field regime $\Lambda r^2 \gg \sqrt{m/r}$, this uncertainty will always be far smaller than the second-order term $15\pi m^2/4 b_\mathrm{eff}^2$. Consequently, the present discussion is sufficiently precise for the vast majority of practical applications. The lensing law \emph{does} allow for the existence of extremely small $\Lambda$-dependent terms $O(m^3 \Lambda r^2/b^3_\mathrm{eff})$, such as those claimed by Sereno \cite{Sereno09}; however these effects would clearly be smaller than the implicit ambiguity of unlensed distances. As such, hypothetical corrections $O(m^3 \Lambda r^2/b^3_\mathrm{eff})$ could always be absorbed into the semi-arbitrary definitions of $\{D_L, D_S, D_{LS}\}$ and thus cannot be considered physically meaningful in the present approach.

If we wish to move beyond these limitations, so as to discuss possible corrections $O(m^3 \Lambda r^2/b^3_\mathrm{eff})$ and smaller, we must abandon the use of \emph{unlensed} distances. To this end, we have included an appendix in which our analysis is reformulated in terms of \emph{lensed} angular diameter distances $\{\bar{D}_L, \bar{D}_S, \bar{D}_{LS}\}$. This approach is significantly more technical than the unlensed one, contending as it must with the singularity at $r=0$ and a variety of other complications. The calculation allows us to improve our lensing law (\ref{newalpha}), yielding
\begin{align}\label{newalphalensed}
\alpha_\mathrm{eff} =  \frac{4m}{b_\mathrm{eff}}+\frac{15\pi m^2}{4b_\mathrm{eff}^2} + O(m^3/b_\mathrm{eff}^3),
\end{align}
in precise agreement with the na\"ive formula (\ref{oldalpha}). This ``lensed distance'' result has two key advantages over the unlensed formulation. First, the error term is now independent of $\Lambda$, so we can understand the $\Lambda=0$ lensing law (\ref{oldalpha}) as the $\Lambda\to0$ limit of the general case (\ref{newalphalensed}). Second, should $\Lambda$-dependent terms appear at next order in (\ref{newalphalensed}), we would need to take them seriously: we could not dismiss these contributions as artefacts of a particular unlensed distance convention. The task of extending equation (\ref{newalphalensed}) to higher order lies beyond the scope of this article. Nonetheless, this method (using \emph{lensed} distances) could potentially establish a physically meaningful $\Lambda$--lensing effect, albeit one of very small magnitude.

Moving on from third-order terms, it is also worth recognising that Schwarzschild--de Sitter is an \emph{extreme case}, in which the effects of $\Lambda$ should be particularly pronounced. Consequently, our conclusions are not limited to this specific spacetime. Within a more realistic model (e.g.\ a spatially dispersed lens in $\Lambda$CDM) the cosmological constant can be expected to have an \emph{even weaker} effect on gravitational lensing. In this sense, SdS has served as an effective \emph{litmus test}, placing an informal upper bound on the magnitude of the $\Lambda$--lensing effect in general. Even amongst papers claiming that $\Lambda$ significantly modifies lensing, there is still agreement that the effect will be greatest for Schwarzschild--de Sitter. For instance, take Sch\"ucker's treatment \cite{Schucker09b} of a semirealistic lensing model: the Einstein-Straus spacetime (an FRW cosmology with an SdS vacuole). Here, Sch\"ucker finds a \emph{screening effect}, attenuating the $\Lambda$--lensing interaction relative to his treatment of SdS \cite{Schucker09a}.

Finally, it is worth noting that although our derivations have focused exclusively on the collinear case (wherein the source $S$ lies on the optical axis defined by the lens $L$ and observer $O$) our results are easily generalised beyond this restriction. Suppose we displace $S$ by an infinitesimal distance $\ud X_S$ perpendicular to the optical axis: the Einstein ring image of $S$ at $O$ will then split into two distinct images, one observed above the axis at an angle $\vartheta_\mathrm{obs}+ \ud \vartheta_\mathrm{obs}$, and the other below the axis at $-\vartheta_\mathrm{obs}+\ud \vartheta_\mathrm{obs}$. Using the lensed distance formalism of the appendix, we see that $\ud X_S= 2\bar{D}_S \ud \vartheta_\mathrm{obs}$; this relation holds true for the accurate model (\ref{lensedDs}) and also follows from the na\"ive picture (figure \ref{lensedtrianglesfig}). As such, the predictions of the two approaches remain in agreement even when the source lies off the optical axis.

\begin{acknowledgments}
The author is supported by a research fellowship from the Royal Commission for the Exhibition of 1851. He thanks John Peacock for useful discussions, and for bringing this problem to his attention.
\end{acknowledgments}

\appendix*
\section{Lensed Distances}\label{lensed}
\begin{figure}
\includegraphics[scale=.45]{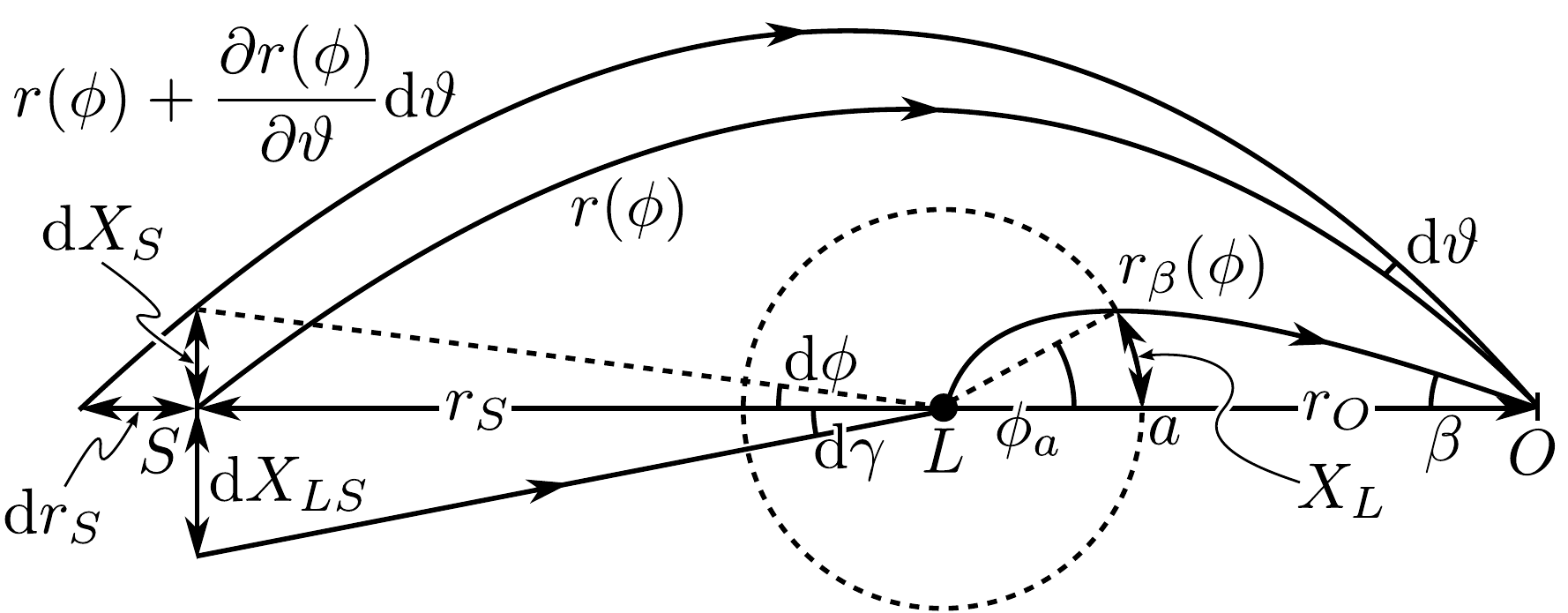}
\caption{Lensed angular diameter distances $\{\bar{D}_L, \bar{D}_S, \bar{D}_{LS}\}$ are defined by the  physical lengths $\{X_L, \ud X_S, \ud X_{LS}\}$ prescribed by the curved light paths of the SdS spacetime (\ref{SdS}). Due to the singularity at $L$, we are forced to construct $X_L$ at some radius $a>0$ and take the limit $a\to0$ at the end of the calculation.}\label{lensedDfig}
\centering
\end{figure}

We define lensed angular diameter distances by
\begin{align}\label{lensedDs}\bs
\bar{D}_{L} &\equiv \lim_{a\to 0}\left\{\left.\frac{\ud X_L}{\ud \beta_\mathrm{obs}}\right|_{\beta=0}\right\},\\
\bar{D}_{S} &\equiv \frac{1}{2}\frac{\ud X_S}{\ud \vartheta_\mathrm{obs}},
\\
\bar{D}_{LS} &\equiv \frac{\ud X_{LS}}{\ud \gamma_\mathrm{coord}},\es
\end{align}
where $\{X_L, \ud X_S, \ud X_{LS}\}$ are the physical lengths constructed in figure \ref{lensedDfig}.\footnote{Alternatively, one could try to construct $\{\bar{D}_{L}, \bar{D}_{S}, \bar{D}_{LS}\}$  using the ratios of infinitesimal \emph{solid} angles and physical \emph{areas}. This approach agrees with (\ref{lensedDs}) for $\{\bar{D}_{L},\bar{D}_{LS}\}$ by virtue of the rotational symmetry about $SO$; however, it fails to correctly define $\bar{D}_S$. To see this, consider a small sphere of radius $\rho$ centred on $S$: this object presents a physical area $\delta A=\pi \rho^2$ and is observed at $O$ as an Einstein ring with solid angle $\delta\Omega= (2\pi \vartheta_\mathrm{obs})\times 2\rho  (\ud \vartheta_\mathrm{obs}/\ud X_S)$. The solid angle approach would therefore define $(\bar{D}_S)^{2} \equiv\lim_{\rho\to0}\delta A/\delta\Omega=0$, which is clearly unphysical. This issue stems from the infinite magnification of $S$ along directions tangential to the Einstein ring; in order to obtain a nonzero angular diameter distance, we must restrict our attention to the behaviour of light along radial slices of the ring, leading us back to $\bar{D}_S$ as defined in (\ref{lensedDs}).} Each of these definitions has a technical quirk as follows:
\begin{itemize}
\item[$\bar{D}_{L}$] Due to the singularity at $L$, we are forced to construct $X_L$ at a some radius $a>0$. Consequently, the limit $a\to0$ is included in the definition. 
\item[$\bar{D}_{S}$] Within the $(r,\phi)$ plane, there are two images of $\ud X_S$ at $O$: one from light which travels above $L$ (as pictured in figure \ref{lensedDfig}) and one from light which travels below $L$ (not pictured). By symmetry, these images will have the same angular size $\ud \vartheta_\mathrm{obs}$ to first order in $\ud X_S$. Dividing $\ud X_S$ by the sum of the angular sizes of the two images generates the  factor of $1/2$ in the definition above. As we will see, this protocol (summing the images) is necessary for $\bar{D}_S$ to be consistent with the unlensed $D_S$ in the $m\to0$ limit.
\item[$\bar{D}_{LS}$] There can be no observers at the singularity, so it is not immediately obvious how one should define an observed angle $\ud \gamma$ at $L$. Fortunately, the spherical symmetry of the spacetime motivates a natural way to proceed: we consider small circles centred on $L$ and evaluate the fraction of their arclength that is subtended. A moment's thought confirms that this definition is equivalent to the coordinate angle $\ud \gamma_\mathrm{coord}$, hence its appearance in the definition.
\end{itemize}
With these definitions at hand, we have two tasks ahead of us: first, to evaluate $\{\bar{D}_L, \bar{D}_S, \bar{D}_{LS}\}$ in terms of $r_O$ and $r_S$; second, to recast the synthesis of section \ref{Synth} in terms of lensed distances.

\subsection{Evaluation}
Let us begin our calculations with $\bar{D}_{L}$. Note that
\begin{align}\nonumber
\ud \beta_\mathrm{obs}&= \sqrt{\frac{1+v}{1-v}}f^{1/2}_O \ud \beta_\mathrm{coord}\\
&=  (1+\sqrt{1-f_O}) \ud \beta_\mathrm{coord}, 
\end{align}
as follows from the discussion of aberration in section \ref{Physangles}. Hence the definition (\ref{lensedDs}) can be written as
\begin{align}\label{toDL'}
\bar{D}_{L} &\equiv \frac{1}{1+\sqrt{1-f_O}} \lim_{a\to 0}\left\{\left.\frac{\ud X_L}{\ud \beta_\mathrm{coord}}\right|_{\beta=0}\right\}.
\end{align}
To evaluate the derivative, we will need to determine the light path $r_\beta(\phi)$ that defines $X_L$ in figure \ref{lensedDfig}. Note that this path is \emph{not} well described by (\ref{lightpath}): we will be taking the limit $\beta\to0$ wherein light crosses the event horizon and $m/R\to \infty$. Instead, we shall consider the tangent vector $k^\mu =(k^t, k^r,0,k^\phi)$ of the (affinely parametrised) null geodesic that the light beam describes. Using $k^\mu k_\mu=0$, and noting that the path is future directed, outward and clockwise ($k^t>0$, $k^r >0$, $k^\phi <0$) we find
\begin{align}\nonumber
r'_\beta\equiv\frac{\ud r_\beta}{\ud \phi}= \frac{k^r}{k^\phi}&= \frac{\sqrt{(k_t)^2-r_\beta^{-2}(k_{\phi})^2 f(r_\beta)}}{r_\beta^{-2}k_\phi}\\\label{drdphi}
&=-r_\beta \sqrt{r_\beta^2 \mu^2 - f(r_\beta)},
\end{align}
where $\mu \equiv k_t/k_\phi$ is a constant due to the spherical symmetry and time invariance of the SdS metric (\ref{SdS}). We can determine $\mu^2$ by considering the angle formed at $O$: 
\begin{align}
\tan(\beta_\mathrm{coord})=\frac{r_\beta}{|r'_\beta|} = \frac{1}{\sqrt{r_O^2 \mu^2 - f_O}},
\end{align}
where the first equality follows by same logic as lead to (\ref{thetas}). Solving this for $\mu^2$, and inserting the result into (\ref{drdphi}), we obtain
\begin{align}
\!\frac{\ud r_\beta}{\ud \phi}= -r_\beta \sqrt{\left(\frac{r_\beta}{r_O}\right)^2\!\left(\cot^2(\beta_\mathrm{coord})+f_O\right) -f(r_\beta)}.
\end{align}
This can be integrated as follows:
\begin{align}\nonumber
\phi_a &=\int^{a}_{r_O}\frac{\ud \phi}{\ud r_\beta} \ud r_\beta\\
&=  \int^{r_O}_{a}\frac{\ud r}{r \sqrt{(r/r_O)^2(\cot^2(\beta_\mathrm{coord})+f_O) -f(r)}},
\end{align}
where $\phi_a$ is defined in  figure \ref{lensedDfig}. Finally, note that $X_L = a \phi_a$ implies
\begin{align}\nonumber
\left.\frac{\ud X_L}{\ud \beta_\mathrm{coord}}\right|_{\beta=0}=a \left.\frac{\ud \phi_a}{\ud \beta_\mathrm{coord}}\right|_{\beta=0}&= a \int^{r_O}_{a}\frac{r_O\ud r}{r^2}\\
&=r_O-a,
\end{align}
and hence (\ref{toDL'}) becomes
\begin{align}\label{D'L}
\bar{D}_{L}=  \frac{r_O}{1+\sqrt{1-f_O}}.
\end{align}
As one would expect, this lensed distance agrees with the corresponding unlensed distance (\ref{Ds}) in the $m\to0$ limit.

Turning our attention to $\bar{D}_S$, we see from figure \ref{lensedDfig} that $\ud X_S$ is defined by two curved lines: the light path $r(\phi)$, described by (\ref{lightpath}), and the nearby light path 
\begin{align}
r(\phi) + \frac{\partial r(\phi)}{\partial \vartheta_\mathrm{obs}}\ud \vartheta_\mathrm{obs},
\end{align}
where the partial derivative is such that $\{r_O, m, \Lambda\}$ are held constant. 
Consequently, 
\begin{align}\nonumber
\ud r_S &= \left[r(\phi) +\frac{\partial r(\phi)}{\partial \vartheta_\mathrm{obs}}\ud \vartheta_\mathrm{obs}\right]_{\phi=\phi_S} - r_S\\\label{drS}
&= \left(\frac{\partial \vartheta_\mathrm{obs}}{\partial r_S}\right)^{-1}\ud \vartheta_\mathrm{obs},
\end{align}
and hence
\begin{align}\label{dXS}
\ud X_S = r_S \ud \phi = \frac{r_S}{r'_S} \ud r_S =  \frac{r_S}{r'_S} \left(\frac{\partial \vartheta_\mathrm{obs}}{\partial r_S}\right)^{-1}\ud \vartheta_\mathrm{obs},
\end{align}
where 
\begin{align}
r'_S\equiv \left.\frac{\ud r(\phi)}{\ud\phi}\right|_{\phi=\phi_S}.
\end{align}
Inserting (\ref{dXS}) into the definition (\ref{lensedDs}), we arrive at
\begin{align}
\bar{D}_S=\frac{1}{2} \frac{r_S}{r'_S}  \left(\frac{\partial \vartheta_\mathrm{obs}}{\partial r_S}\right)^{-1},
\end{align}
the first factor of which can be obtained in the same fashion as equation (\ref{r/r'}),
\begin{align}\label{r/r'S}
\frac{ r_S}{r'_S}=  \frac{2m^{1/2}}{r_S\sqrt{r_O^{-1}+ r_S^{-1}}} + \frac{15\pi m}{32 r_S} + O(m^{3/2}/r^{3/2}),
\end{align}
and the second factor of which follows by differentiation of equation (\ref{thetaobs}):
\begin{align}
\frac{\partial \vartheta_\mathrm{obs}}{\partial r_S}= \frac{m^{1/2}\left(1+\sqrt{1-f_O}\right)}{r_Or_S^2 \left(r_O^{-1}+r_S^{-1}\right)^{3/2}}+ O(m^{3/2}/r^{5/2}).
\end{align}
Thus
\begin{align}\nonumber
\bar{D}_S& = \frac{r_O+r_S}{1+\sqrt{1-f_O}}\left(1+ \frac{15\pi m^{1/2}}{64}\sqrt{r_O^{-1}+r_S^{-1}}  \right)\\\label{D'S}
&\quad {}+O(m).
\end{align}
Note that the $m\to 0$ limit of $\bar{D}_S$ agrees with the unlensed distance (\ref{Ds}); this would not have been the case had we neglected the second image of $S$ at $O$ and the associated factor of $1/2$ in the definition (\ref{lensedDs}). 

Finally, $\bar{D}_{LS}$ is trivial: the defining light path is radial, so $\ud X_{LS} = r_S \ud\gamma_\mathrm{coord}$ and hence
\begin{align}\label{D'LS}
\bar{D}_{LS}= r_S.
\end{align}
This is in exact agreement with the unlensed value (\ref{Ds}) assigned by the area-preserving convention.

\subsection{Application}
\begin{figure}
\includegraphics[scale=.45]{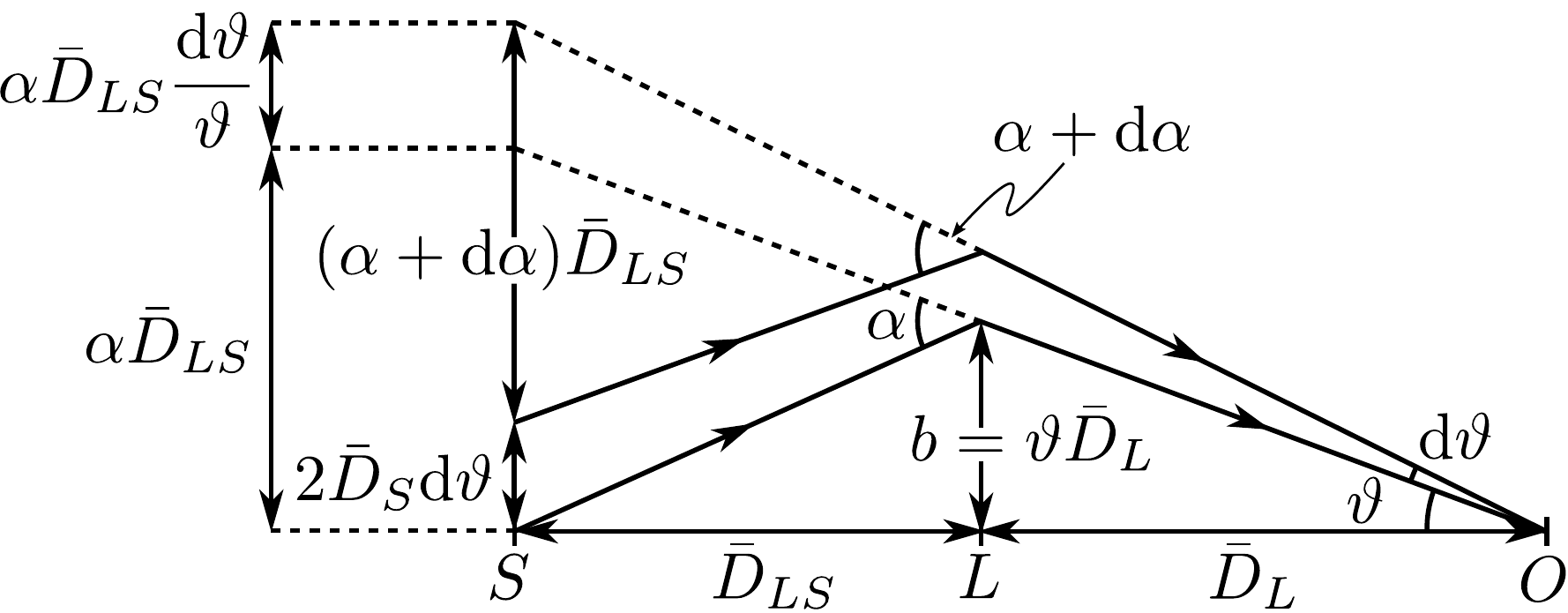}
\caption{The schematic representation of gravitational lensing that replaces figure \ref{triangles} when one exclusively uses \emph{lensed} angular diameter distances $\{\bar{D}_L,\bar{D}_S,\bar{D}_{LS}\}$. Once again, this picture makes use of the small angle approximation.}\label{lensedtrianglesfig}
\centering
\end{figure}

In order to adapt the logic of section \ref{Synth} to the case at hand, we must first replace  figure \ref{triangles} with a representation of the lensing geometry in terms of lensed angular diameter distances: this is achieved with figure \ref{lensedtrianglesfig}. As the image makes clear, we can identify the impact parameter 
\begin{align}\label{blensed}
b=\vartheta \bar{D}_L,
\end{align}
much as before; however, the role of the bending angle $\alpha$ is a little more obscure. To explore this aspect, we equate the two lengths on the left of figure \ref{lensedtrianglesfig}:
\begin{align}\label{equallengths}
\alpha \bar{D}_{LS}\left(1 + \ud\vartheta/\vartheta\right)= (\alpha + \ud \alpha) \bar{D}_{LS} + 2 \bar{D}_S \ud \vartheta.
\end{align}
Noting that $\ud \alpha= (\ud \alpha/\ud b)\ud b=  (\ud \alpha/\ud b) \bar{D}_{L}\ud \vartheta$, equation (\ref{equallengths}) becomes
\begin{align}\label{alphalensed}
\left(1-b\frac{\ud}{\ud b}\right) \alpha= \frac{2 \vartheta \bar{D}_S}{\bar{D}_{LS}},
\end{align}
which we interpret as a differential equation for $\alpha=\alpha(b)$.

We now apply the reasoning of section \ref{Synth} and use equations (\ref{blensed}) and (\ref{alphalensed}) to \emph{define} effective values of the impact parameter and bending angle in the absence of asymptotic flatness. Recalling that figure \ref{lensedtrianglesfig} invoked the small angle approximation, we have
\begin{align}\bs\label{beffdeflensed}
b_\mathrm{eff}&\equiv \vartheta_\mathrm{obs} \bar{D}_L + O(\vartheta_\mathrm{obs}^3 \bar{D}),\es
\end{align}
and define $\alpha_\mathrm{eff}$ as the solution to
\begin{align}\label{alphaeffdeflensed}
\left(1-b_{\mathrm{eff}}\frac{\ud}{\ud b_\mathrm{eff}}\right) \alpha_\mathrm{eff}= \frac{2\vartheta_\mathrm{obs}\bar{D}_S }{\bar{D}_{LS}} + O(\vartheta_\mathrm{obs}^3 ),
\end{align}
such that $\alpha_\mathrm{eff}\to 0$ as $b_\mathrm{eff}\to \infty$. These definitions replace (\ref{effdef}) when one describes the lensing geometry in terms of lensed distances.

Finally, we insert equations (\ref{thetaobs}), (\ref{D'L}), (\ref{D'S}), and (\ref{D'LS}) into the definitions (\ref{beffdeflensed}) and (\ref{alphaeffdeflensed}):
\begin{align}\label{befflensed}
b_{\mathrm{eff}}&=\frac{2m^{1/2}}{\sqrt{r_O^{-1} + r_S^{-1}}} + \frac{15 \pi m}{32} + O\left(\vartheta^3_\mathrm{obs}r\right),
\end{align}
and
\begin{align}\nonumber
\bigg(1{}- {}&b_\mathrm{eff}\frac{\ud}{\ud b_\mathrm{eff}}\bigg)\alpha_{\mathrm{eff}}\\\nonumber
&= 4m^{1/2} \sqrt{r_O^{-1} + r_S^{-1}}\left(1+ \frac{15\pi m^{1/2}}{32}\sqrt{r_O^{-1} + r_S^{-1}}\right)\\\label{alphaefflensed}
&\quad{}+ O\left(\vartheta^3_\mathrm{obs}\right).
\end{align}
Rearranging (\ref{befflensed}) gives
\begin{align}\nonumber
\!\!\sqrt{r_O^{-1} + r_S^{-1}}&=\frac{2m^{1/2}}{b_{\mathrm{eff}}-  (15 \pi m/32) + O\left(\vartheta^3_\mathrm{obs}r\right)}\\
&= \frac{2m^{1/2}}{b_\mathrm{eff}}+ \frac{15 \pi m^{3/2}}{16 b^2_\mathrm{eff}} + O\!\left(\frac{m^{5/2}}{b_\mathrm{eff}^{3}}\right),
\end{align}
which we insert into (\ref{alphaefflensed}) to get
\begin{align}
\bigg(1- b_\mathrm{eff}\frac{\ud}{\ud b_\mathrm{eff}}\bigg)\alpha_{\mathrm{eff}}= \frac{8m}{b_\mathrm{eff}}+ \frac{45 \pi m^2}{4b_\mathrm{eff}^2} + O(m^3/b_\mathrm{eff}^3).
\end{align}
The unique solution to this equation (such that $\alpha_\mathrm{eff}\to 0$ as $b_\mathrm{eff}\to \infty$) is then
\begin{align}
\alpha_\mathrm{eff} =  \frac{4m}{b_\mathrm{eff}}+\frac{15\pi m^2}{4b_\mathrm{eff}^2} + O(m^3/b_\mathrm{eff}^3),
\end{align}
which precisely replicates the na\"ive formula (\ref{oldalpha}) without a $\Lambda$-dependent error term.

\bibliography{LambdaLens}
\end{document}